\documentstyle[12pt]{article}
\input{epsf}

\input{epsf}
\input{rotate}
\begin{document}
\title{\bf{BIG--BANG NUCLEOSYNTHESIS AS A PROBE OF THE GEOMETRY OF
	SUPERSPACE}}
\author{
        A. Llorente and J. P\'erez-Mercader\\
      \normalsize \it  Laboratorio de Astrof\'\i sica Espacial y
	F\'\i sica Fundamental \\
      \normalsize \it	Apartado 50727 \\
	\normalsize \it 28080 Madrid (Spain)
        }
\date{}
\maketitle
\vskip -110mm
\hglue 100mm LAEFF-94/15
\vskip 109mm

\begin{abstract}
We study the effect from a general
ultralocal supermetric on primordial nucleosynthesis for
Friedmann-Robertson-Walker Cosmology.
The parameter $\lambda$ of the supermetric changes the effective
number of degrees of
freedom, $g_*$, which modifies the Friedmann
equation. This modification produces variations in the production of
primordial $^4He$.
The observations of the primordial abundances
of
light elements ($^4He$, $D$, $^3He$ and $^7Li$) allow to estimate
bounds on the values of $\lambda$
consistent with these observations. We obtain $0.87 \le \lambda \le
1.04$. In addition we analyze the importance of $\lambda < 1$
to explain possible incompatibilities in the standard Big-Bang
nucleosynthesis.
\end{abstract}

\pagebreak

In the first quantization of gravity \cite{dw} one needs to
introduce
a ``metric
of metrics", the supermetric $G_{ijkl}$. The most general ultralocal
supermetric depends on a free parameter $\lambda$, which is usually fixed,
on arbitrary grounds, to $\lambda=1$. For second quantized
gravity, the value of $\lambda$ appears, e.g., on
the $\beta$-functions for Newton's Constant and the Cosmological
Constant \cite{juan}, the two dimensionful parameters in the
four dimensional gravitational action.

In this paper, we study the effect of $\lambda$ on cosmological light
element
production and show how the observational values of the
abundances of $^4He$, $D$,
$^3He$ and $^7Li$ can be used to bound $\lambda$.

Let us consider General Relativity (GR) in its 3+1 ADM formulation
\cite{adm}. The hamiltonian $H$ is
\begin{equation}
H=\int d^3x(NH_G+N_iH^i)
\end{equation}
where the hamiltonian constraint is
\begin{equation}
H_G\equiv 16\pi G G_{ijkl}\pi^{ij}\pi^{kl}-{\sqrt{h}\,\,^3R \over
16\pi G},
\end{equation}
with the supermetric given by
\begin{equation}
G_{ijkl}\equiv {1 \over
2\sqrt{h}}\left(h_{ik}h_{jl}+h_{il}h_{jk}-h_{ij}h_{kl} \right)
\end{equation}
and where $h_{ij}$ is the induced spatial metric, $N$ is the lapse
function and $N^i$ the shift vector. $G$ is Newton's Constant.
The $\pi^{ij}$ are the momenta conjugate to $h_{ij}$.

In the presence of matter the constraint equation, $H_G=0$, becomes
\begin{equation}
16\pi G G_{ijkl}\pi^{ij}\pi^{kl}-{\sqrt{h} ^3R \over 16\pi G} - T =0
\label{wdw}
\end{equation}
where $T$ is the ${T^0}_0$ component of the stress-energy tensor of the
matter fields.

For a Friedmann--Robertson--Walker metric
\begin{equation}
ds^2=dt^2-a^2(t) \left\{ {dr^2 \over 1-kr^2}+r^2d\theta ^2+r^2 sin ^2
\theta d\phi ^2 \right\}\, ,
\label{rw}
\end{equation}
Eq.(\ref{wdw}) is nothing but the familiar Friedmann equation
\begin{equation}
H_G=0 \to \left( {\dot a\over a}\right) ^2 + {k\over a^2}=
{8\pi G \over 3}\rho \, .
\label{f}
\end{equation}

As shown by deWitt in Ref.\cite{dw}, the most general ultralocal
supermetric is given by
\begin{equation}
G_{ijkl}\equiv {1 \over
2\sqrt{h}}\left(h_{ik}h_{jl}+h_{il}h_{jk}-\lambda h_{ij}h_{kl} \right) \, .
\label{sm}
\end{equation}

The quantity $\lambda$ parametrizes the distance between me\-trics in
superspace and, in principle, may take any real value except
$\lambda={2 \over 3}$ for which the supermetric is singular.

When using (\ref{sm}) in Equation (\ref{wdw}), and inserting a homogeneous
and
isotropic metric as in (\ref{rw}), one easily finds the following
{\it modified} Friedmann equation
\begin{equation}
{1 \over 3\lambda -2 }\left( {\dot a\over a}\right) ^2 + {k\over a^2}=
{8\pi G \over 3}\rho \, .
\label{fl}
\end{equation}

This equation shows how the geometry of superspace percolates into
the physics of the metric (\ref{rw}). As far as the geometry of superspace
is concerned, there is no reason to have one or another value of
$\lambda$; but as (\ref{fl}) demonstrates, the parameter $\lambda$ changes
the expansion rate of the Universe and, thus, has
observable cosmological
implications.

In what follows, we will see how the $\lambda$-dependence in
Eq.(\ref{fl})
can be understood as an effective modification to the number of
internal degrees of freedom contributing to the density $\rho$ of the
relativistic gas.

When we do not take superspace into account, the Friedmann equation
for a radiation dominated Universe obtained from Eq.(\ref{f}) is \cite{kt}
\begin{equation}
H^2={{8\pi } \over {3m_p^2}}\rho ={{8\pi } \over
{3m_p^2}}{{\pi ^2} \over {30}}g_*T^4\to H\approx {{1.67}
\over {m_p}}g_*^{1/2}T^2
\label{tem}
\end{equation}
where $g_*$ is defined by
\begin{equation}
g_* \equiv \sum\limits_{i=fermions} {{7 \over 8}g_i\left(
{{{T_i} \over T}} \right)^4}+\sum\limits_{i=bosons}
{g_i\left( {{{T_i} \over T}} \right)^4}
\end{equation}
and $g_i$ counts the number of internal degrees of freedom
of the {\it i}-th species, $T$ is the
photonic temperature and $T_i$ the temperature of the corresponding
{\it i}-th species. When superspace effects are taken into consideration,
we have
seen that Eq. (\ref{tem}) has to be modified according to
Eq.(\ref{fl}), and
therefore Eq.(\ref{tem}) changes into
\begin{equation}
H\approx {{1.67}
\over {m_p}}{\sqrt{3\lambda -2}}g_*^{1/2}T^2\equiv{{1.67}
\over {m_p}} g_*^{1/2}(\lambda)T^2
\label{teml}
\end{equation}
where in the last term we have redefined the effective $g_*$ as
\begin{equation}
g_*(\lambda)=(3\lambda -2)\left[\sum\limits_{i=fermiones}
{{7 \over 8}g_i\left(
{{{T_i} \over T}} \right)^4}+\sum\limits_{i=bosones}
{g_i\left( {{{T_i} \over T}} \right)^4}\right ]
\end{equation}
so that, a general supermetric can be interpreted as inducing a
redefinition of the parameter $g_*$ into an
effective value which depends on $\lambda$.

If we take the number of neutrino species, $N_\nu$, to be 3,
as indicated by the CERN results
\cite{cern},
for temperatures of order of $MeV$, $g_*(\lambda)$ is
\begin{equation}
g_*(\lambda)=(3\lambda -2) \left. \left[ 2+ {7 \over 8}2N_\nu+{7 \over
8}2 \cdot 2 \right] \right|_{N_\nu=3} =(3\lambda -2) \times 10.75\, ,
\label{gl}
\end{equation}
and since $^4He$ production is
sensitive to the value of
$g_*$, a change in $\lambda$
will change the amount of $^4He$ produced in the Big Bang
nucleosynthesis.

We note the following general behavior:
if $\lambda$ is equal to one, the standard results are
recovered; if $\lambda$ is larger than one, the Big Bang
nucleosynthesis produces {\it more} primordial $^4He$, and if $\lambda$ is
smaller than one the amount of $^4He$ produced will be less than in the
standard case.

Before we use the observational data on primordial abundances to place
bounds on
$\lambda$ via Eq.(\ref{teml}), we briefly discuss the available data
on
$^4He$ \cite{olive94}, $D$, $^3He$ and $^7Li$ \,\cite{wea}.

It is possible to fit the observed $^4He$ abundance with respect to
the abundances of $O$ and $N$ in $H_{II}$-regions. The primordial
$^4He$ corresponds to zero metallicity. A range
consistent with observations is \cite{olive94}
\begin{equation}
Y_P=0.232 \pm 0.003 \pm 0.005\, ,
\end{equation}
where the first error is $1\sigma$ statistical, and the second error is
systematic (estimated at 2$\%$).

Through solar reactions, the initial amount (pre-solar) of $D$ is
burned to form
$^3He$, while the
initial amount of $^3He$ is kept approximately constant in time.
Therefore, the present $^3He$ abundance will be an indicator of
the initial pre-solar, combined, $D$+$^3He$ abundance.
The former is measured
in gas-rich meteorites \cite{grm}, in the solar wind \cite{sw}
and in the lunar soil
\cite{ls}, giving a value
\begin{equation}
3.3 \times 10^{-5} \le \left. {N(^3He) \over N(H)}\right|_
{grm} \left( \approx \left. {N(^3He+D) \over N(H)}\right|_
{pre-solar}\right)\le 4.9 \times 10^{-5}.
\end{equation}

Obtaining primordial abundances from these pre-solar
abundances is model-dependent, although it is possible to infer an
upper bound on the primordial abundances of $^3He$+$D$. This bound is
\cite{7Li1,7Li2}
\begin{equation}
\left.{N(^3He+D) \over N(H)}\right|_P\le 10^{-4}.
\end{equation}

Furthermore, since carbonaceous chondrites (the oldest meteorites) are
believed to provide a sample of pre-solar abundances, the
fraction of $^3He$ measured in them gives us an estimate of
pre-solar abundances of $^3He$ (without the solar contamination
due to deuterium burning). These values are \cite{cc1,cc2}
\begin{equation}
1.3\times 10^{-5} \le \left.{N(^3He) \over
N(H)}\right|_{cc}\left(\approx
\left.{N(^3He) \over N(H)}\right|_{pre-solar}\right) \le 1.8 \times
10^{-5}.
\end{equation}

The difference between the pre-solar abundance of $D$+$^3He$ and
the pre-solar abundance of $^3He$ will be the pre-solar abundance
of deuterium, that is,
\begin{equation}
1.8 \times 10^{-5} \le \left. {N(D) \over N(H)} \right| _{pre-solar} \le 3.3
\times
10^{-5},
\end{equation}
and since deuterium is destroyed but is not produced (in significant
quantities) in stellar processes, the minimum amount
of deuterium in the pre-solar system
gives us a lower bound on primordial deuterium. Thus
\begin{equation}
\left.{N(D) \over N(H)}\right|_P \ge \left.{N(D) \over N(H)}\right|_{pre-solar}
\ge 1.8 \times
10^{-5}.
\end{equation}

Finally, for $^7Li$, very old Population II stars (halo stars) can
provide an estimate
of primordial
$^7Li$ abundance. This hypothesis is supported by
two observational facts: the plot of observational abundances of $^7Li$
versus iron metallicity has a plateau for low metallicity, implying
that metal-poor stars do not produce significant amounts of $^7Li$;
on the other hand,
the plot of observational abundances of $^7Li$
versus effective temperature has another plateau for high temperatures,
which implies that stars with high temperatures do not destroy their
initial $^7Li$.

For
the hotter and more metal-poor Population-II stars in the halo,
one has \cite{wea}
\begin{equation}
\left. {N(^7Li)\over N(H)} \right| _P\approx \left. {N(^7Li)\over N(H)}
\right| _{II-Pop}=1.2 \pm 0.2 \times 10^{-10},
\end{equation}
so that an upper bound for primordial $^7Li$ is
\begin{equation}
\left.{N(^7Li)\over N(H)}\right|_P \le 1.4 \times 10^{-10}.
\end{equation}

We summarize the above discussion in Table 1, where we also list
the allowed values for $\eta_{10}$ obtained by comparing the observational
data with theoretical standard nucleosynthesis; the
neutron life-time $\tau_{1/2}$ has been taken between
10.19 and 10.35 minutes
\footnote{An accurate experiment to determine $\tau_{1/2}$ is
described, e.g., in Ref.\cite{neutron}.} and
$N_\nu = 3$. Notice that
\begin{equation}
2.8 \le \eta_{10} \le 4.0
\end{equation}
is compatible
\footnote{This can also be expressed in terms of the
combination $\Omega_B h_{100}^2$, where $\Omega_B$ is the baryonic
fraction of the critical density and $h_{100}$ is the normalized
Hubble parameter, as
\begin{equation}
0.010 \le \Omega_B h_{100}^2 \le 0.015\, .
\end{equation}
}
with all the observational data.

We are now ready to derive bounds on $\lambda$ from the observed
abundances of the light elements. Since the theoretical primordial
mass fraction of $^4He$ is well fitted over
the range $2.5\le \eta_{10} \le 10$ by (Ref. \cite{oea}) the equation
\begin{equation}
Y_P=0.228+0.010ln(\eta_{10})+0.012(N_\nu-3)+0.017(\tau_{
1/2}-10.27 min)
\label{yp}
\end{equation}
for temperatures of order {\it MeV}, we obtain the following equation
by using Eq.(\ref{gl})
in (\ref{yp})
to write $N_\nu$ as a function of $\lambda$ and then solving for
$\lambda$,
\begin{equation}
\lambda=1+4.52\left[(Y_P-0.228)-0.010ln\eta_{10}-0.017(\tau_{
1/2}-10.27)\right] \, .
\end{equation}
Here $\tau_{1/2}$ must be given in minutes.

In Fig.1, we have plotted $\lambda$ versus $\eta_{10}$ (or,
equivalently,
$\Omega_Bh_{100}^2$), with $Y_P$ and $\tau_{1/2}$ as parametric
variables. As $\eta_{10}$ is increased, the parameter  $\lambda$
decreases slowly since $\lambda$ evolves as a logarithm with
$\eta_{10}$. The change in $\lambda$ with neutron lifetime is also
small for the allowed band in $\tau_{1/2}$.

The most pronounced effect of $\lambda$ shows up in the $^4He$
abundance. When $\lambda$ is increased, the amount of $^4He$ also
increases. Therefore $\lambda$ can substantially disturb the Big-Bang
nucleosynthesis\footnote{We are going to analyze the influence
of $\lambda$ only on $^4He$ production. It can be seen that the effect of
$\lambda$ on
the remaining light elements is negligible compared with observational
errors.}. This can be used in two different ways: the observations
of primordial $^4He$ give bounds on the values of $\lambda$ or, if the
observational data on primordial $^4He$ are too
low, a value of $\lambda$ smaller than 1 helps to bring back
consistence into Big-Bang nucleosynthesis.

Concerning the first point,
we could let $Y_P$, $\tau_{1/2}$ and
$\eta_{10}$ vary inside their bounds, to give a range for $g_*$ and
from there deduce upper and lower bounds on $\lambda$. For the
extreme values quoted in Table 1,
\begin{equation}
0.221 \le Y_P \le 0.243
\end{equation}
\begin{equation}
10.19\,\,\, min \le \tau_{1/2} \le 10.35\,\,\, min
\end{equation}
\begin{equation}
2.8 \le \eta_{10} \le 4.0
\end{equation}
Using equation (\ref{yp}), one finds that
\begin{equation}
7.5 \le g_* \le 11.6 \,\,\,\,\,\,\,\,(1.1 \le N_\nu \le 3.5).
\end{equation}

If we consider the particle contents in the standard model, these bounds
over
$g_*$ imply
\begin{equation}
7.5 \le (3\lambda -2) \cdot 10.75 \le 11.6 \, ,
\end{equation}
which translates into the following range for $\lambda$,
\begin{equation}
0.90 \le \lambda \le 1.03 \, .
\label{cl2}
\end{equation}

The bounds on $\lambda$ depend on the ``choice" of observational
bounds on $Y_P$, $\tau_{1/2}$ and $\eta_{10}$. To see this, we select
less restrictive values for $Y_P$ \cite{wea}: $0.215 \le Y_P
\le 0.245$. The bounds over $\lambda$ now translate into
\footnote{These bounds agree with the
bounds in Ref.\cite{giulini}.}
\begin{equation}
0.87 \le \lambda \le 1.04 \, ,
\label{cl1}
\end{equation}
and therefore the bounds do not change very much for different
observational data, with the values of
$\lambda$ always near 1.

On the other hand,
recent observational and theoretical results on $^4He$ abundances could
be in conflict with the abundances for the remaining light elements,
and thus put in
danger the consistency of the standard Big-Bang nucleosynthesis
\cite{olive94}. A more
accurate theoretical calculation \cite{kernan1,kernan2},
increases the value of $Y_P$ with respect to the results of Ref.
\cite{wea}. Furthermore, the present errors on $\tau_{1/2}$
have been
slightly reduced, and $\tau_{1/2}=10.27 \pm 0.024$ \,\cite{tabla}.
These two effects combine and make slightly
more difficult the agreement between observational and
theoretical values, with the net effect of increasing the discrepancy.

In Fig.2, $Y_P$ is plotted versus $\lambda$, with the previous
considerations already taken into account. The horizontal dashed lines
correspond to the
observations on $Y_P$ \cite{olive94}.
We have plotted the
average $Y_P$ and the upper bounds at
$1\sigma$, $2\sigma$ and $2\sigma+\sigma_{sys}$.

For the average $Y_P$, standard Big-Bang nucleosynthesis
($\lambda=1$, vertical solid line) is {\it incompatible}
with the observations of the other light elements, which fall
between
the two extreme dotted lines ($\eta_{10}=4.0$, $\tau_{1/2}=10.32$ and
$\eta_{10}=2.8$, $\tau_{1/2}=10.22$). The incompatibility is not removed
for observational values of
$Y_P$ inside of the bounds for $1\sigma$--errors or
$2\sigma$--errors.
To recover consistency, systematic errors in
observational data must be considered. However, consistency between
observation and theory is recovered if we consider values of $\lambda$
away from 1. For the average value of
$Y_P$, $Y_P=0.232$, we
obtain $\lambda\approx 0.95$. With
$\lambda \stackrel{<}{_{_{\widetilde{}}}} 0.98$
we are inside of the region of observational errors on $Y_P$ less
than or equal to $1\sigma$, and with $\lambda
\stackrel{<}{_{_{\widetilde{}}}} 0.99$ we are inside of the
region of $2\sigma$ errors. Thus
\begin{equation}
0.93 \le \lambda \le 0.98 \,\,\,\,\,  (1\sigma)
\end{equation}
\begin{equation}
0.91 \le \lambda \le 0.99 \,\,\,\,\,   (2\sigma)
\end{equation}
\begin{equation}
0.89 \le \lambda \le 1.01  \,\,\,\,\,  (2\sigma + \sigma_{sys})
\end{equation}
for errors in observational $Y_P$ of $1\sigma$, $2\sigma$ and
$2\sigma + \sigma_{sys}$ respectively\footnote{These last
values
were obtained
using the $^4He$ production from Ref.\cite{kernan1,kernan2},
whereas
for the previous bounds given in Eq.(\ref{cl2}), \ Ref.\cite{wea}
was used.}.

Therefore with $\lambda <1$ the Big-Bang
nucleosynthesis is
absolutely consistent. Future, more accurate observations, should
clarify the confluence or inconsistence in the observations of
the abundances of all
light elements. If the present situation is confirmed, $\lambda$ {\it could}
be used to explain these discrepancies.

Finally we offer some conclusions. First we
note that there exists a connection between  the physics of superspace
and nucleosynthesis due to the fact that $\lambda$ contributes to the
Friedmann equation in a way which can be interpreted as a modification
to the effective number of degrees of freedom of the
relativistic gas
which dominates the energy density of the Universe during the
Nucleosynthesis era; this has as a consequence that as $\lambda$
increases, so does $Y_P$.

Using the observational data on primordial
$^4He$, $D$, $^3He$ and $^7Li$, we have also obtained bounds on
$\lambda$. These bounds
depend on the values considered for $Y_P$, $\tau_{1/2}$ and
$\eta_{10}$, but they tend to cluster near $\lambda=1$. A
reasonable range of values for $\lambda$ is
\begin{equation}
0.87 \le \lambda \le 1.04 \, .
\end{equation}

{}From the point of view of superspace, therefore,
nu\-cleo\-syn\-the\-sis
implies that the deviation from standard FRW ($\lambda =1$)
is small. But if
these
deviations exist, although small, they could be useful:
as present data indicates, measurements of primordial
$^4He$ could be inconsistent at the $2\sigma$ level with the remaining
observed abundances of $D$, $^3He$ and $^7Li$,
having a slightly low value (if systematic errors are considered,
consistence can be recovered). If the errors in the observations
decrease in
the future, and the value of the $^4He$ abundance was too
small, and
inconsistent
with the abundances of the other light elements, a value of $\lambda$
less
than 1 could be used to understand this disagreement;
this brings with it
a slight deviation from the cosmological standard model
at the time of nucleosynthesis.

On the other hand a value of $\lambda$ different from one, has other
observational cosmological effects, such as modifications on the spectrum
of the density perturbations or on the microwave background anisotropies,
in addition to alterations in the evolution of the angular sizes of luminous
sources with red-shift. These effects will be considered in a separate
paper.

\vskip 8mm

\centerline{\bf ACKNOWLEDGMENTS}

\vskip 5mm

We thank Domenico Giulini for stimulating discussions. This work has
been supported in part by INTA and DGICyT Funds.

\pagebreak

\pagebreak

\begin{figure}
\caption{$\lambda$ versus $\eta_{10}$ (or $\Omega_Bh^2_{100}$). We
have considered the values $Y_P=0.221, 0232, 0.243$. The solid lines
correspond to $\tau_{1/2}=10.27$. The dashed lines are associated to
the
uncertainties in $\tau_{1/2}$: for each $Y_P$, the top dashed line is
$\tau_{1/2}=10.19$ and the bottom dashed line is $\tau_{1/2}=10.35$
[7].
For $Y_P$ fixed, $\lambda$ increases as $\Omega_B$ decreases, and for
$\Omega_B$ fixed, $\lambda$ rises the production of $^4He$. }\label{f1}
\end{figure}

\begin{figure}
\caption{Production of $^4He$ versus $\lambda$. The solid lines
correspond to $\eta_{10}=2.8$ and $\eta_{10}=4.0$ (interval allowed
by observations for the abundances of $^3He$, $D$ and $^7Li$) with
$\tau_{1/2}=10.27$ min. The dotted lines represent the $2\sigma$
limits
on $\tau_{1/2}$ [19]: $\tau_{1/2}=10.22$ and $\tau_{1/2}=10.32$ (for
each $\eta_{10}$). The horizontal dashed lines correspond to the
observational values [6]:
average $Y_P$ and upper bounds $1\sigma$, $2\sigma$ and
$2\sigma+\sigma_{sys}$
($\sigma$ is statistical error and $\sigma_{sys}$ systematic
error). It is clear that the standard Big-Bang nucleosynthesis
($\lambda=1$, vertical solid line) is not compatible inside the
$1\sigma$ and $2\sigma$ errors. It is neccesary
to consider
systematic errors. The compatibility is immediately recovered if we
consider a value of $\lambda$ away from 1.}\label{f2}
\end{figure}
\
\vfil

\setbox0=\hbox{
\begin{tabular}{|c|c|c|c|}
\hline
&&&\\
&{\bf OBSERVATIONAL ABUNDANCES}&{\bf BOUNDS}&{$\eta_{10}$}\\
&&&\\
\hline
&&&\\
$^4He$&$Y_P=0.232 \pm 0.003 \pm 0.005$&$Y_P \le 0.243$&$\eta_{10}
\le 5.1$ \\
&&&\\
\hline
&&&\\
$D+^3He$&$3.3 \times 10^{-5} \le \left.{N(^3He+D) \over N(H)}\right|_
{pre-solar} \le 4.9 \times 10^{-5}$&&\\
&&$\left.{N(^3He+D) \over N(H)}\right|_P\le 1 \times 10^{-4}$&$
\eta_{10} \ge 2.8$\\
$^3He$&$1.3\times 10^{-5} \le \left.{N(^3He) \over N(H)}\right|_{pre-solar} \le
1.8 \times 10^{-5}$&&\\
&&&\\
\hline
&&&\\
$D$&$1.8 \times 10^{-5} \le \left.{N(D) \over N(H)}\right|_{pre-solar} \le 3.3
\times 10^{-5}$&$\left.{N(D) \over N(H)}\right|_P \ge 1.8 \times 10^
{-5}$&$\eta_{10} \le 6.8$\\
&&&\\
\hline
&&&\\
$^7Li$&$\left. {N(^7Li)\over N(H)} \right| _{II-Pop}=1.2 \pm 0.2 \times
10^{-10}$&$\left. {N(^7Li)\over N(H)} \right|_P\le 1.4 \times 10^
{-10}$&$1.6 \le \eta_{10} \le 4.0$\\
&&&\\
\hline
\end{tabular}
}

\begin{table}
\vskip -20mm
\tiny
\rotl 0
\vfill \eject
\caption{
Summary
of the observational data for $^4He$ [6], $D$, $^3He$ and $^7Li$
[7]. By comparing the bounds on these abundances with
theoretical predictions of nucleosynthesis (for
$10.19 \le \tau_{1/2} \le 10.35$ [7] and $N_\nu=3$),
bounds on $\eta_{10}$ are obtained.
For the interval
in $\eta_{10}$ obtained from the $^7Li$
bound, a $40\%$ of systematic error is allowed [7].}
\end{table}

\pagebreak

\begin{figure}
\hbox{
\epsfxsize=140mm
\epsfbox{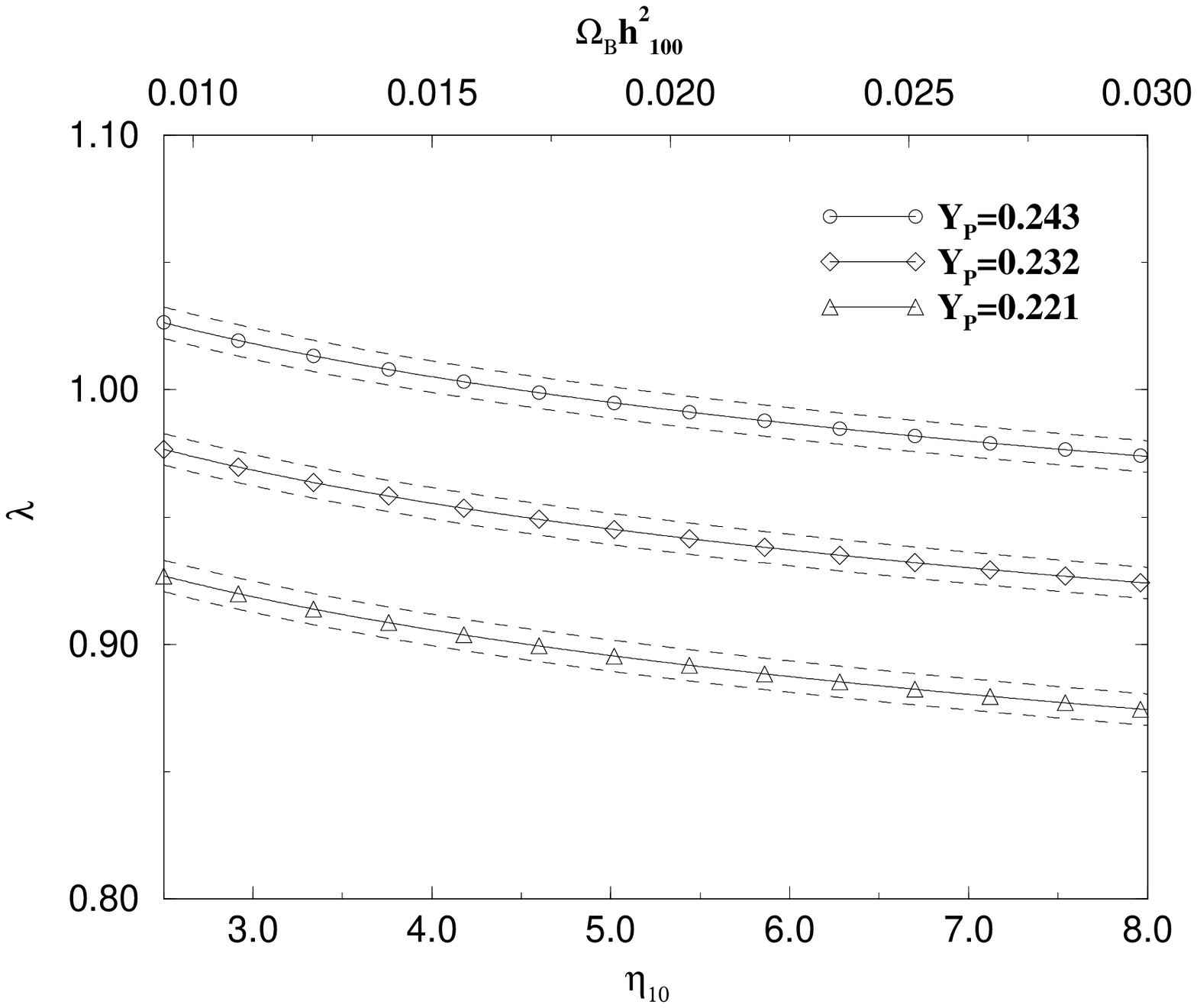}
}
\hbox{\hspace{-20mm}
}
\hfil Fig. \ref{f1} \hfil
\end{figure}

\pagebreak

\begin{figure}
\hbox{
\epsfxsize=140mm
\epsfbox{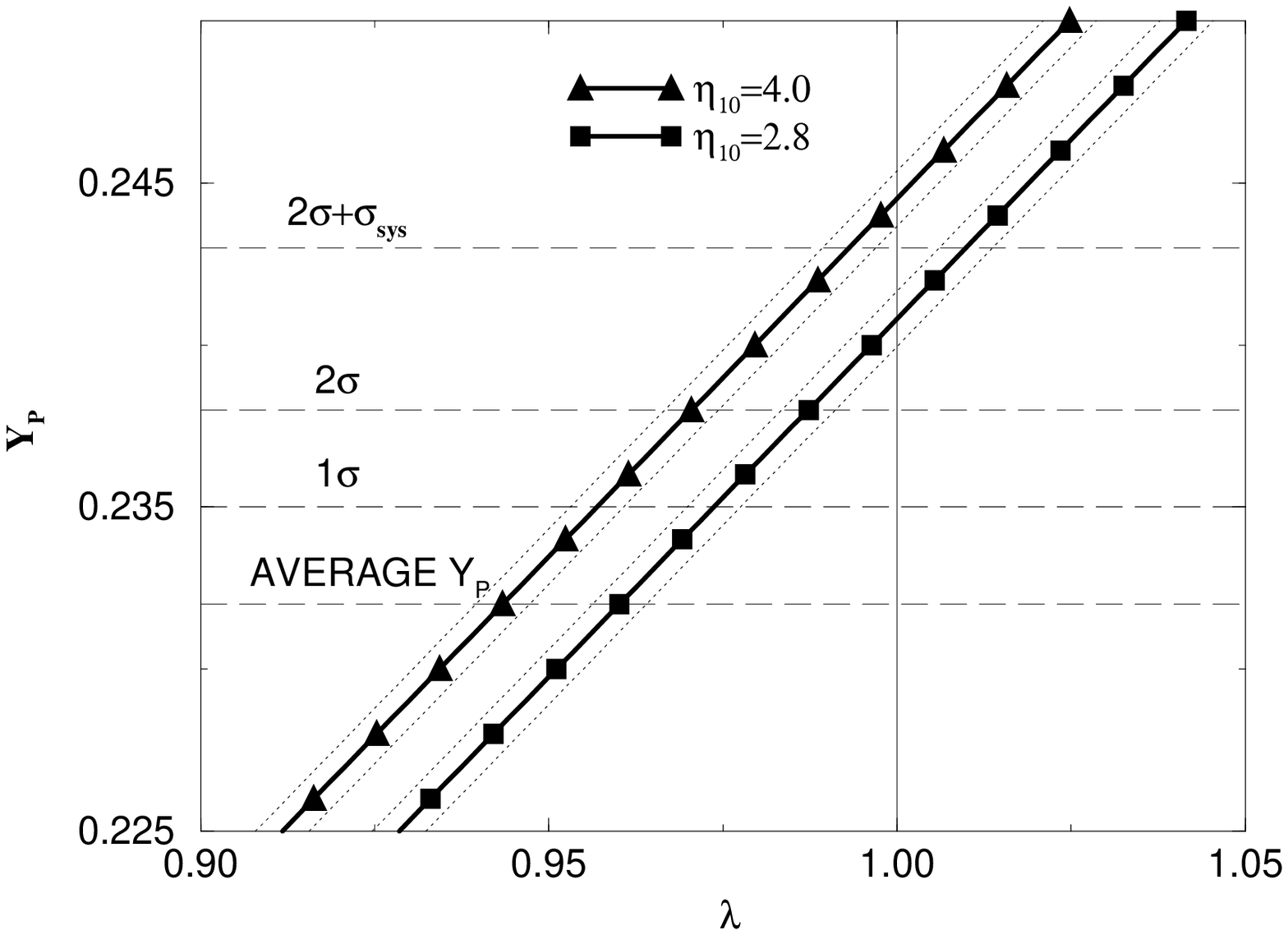}
}
\hbox{\hspace{-20mm}
}
\hfil Fig. \ref{f2} \hfil
\end{figure}


\begin{thebibliography}{20}
\bibitem{dw} B. S. deWitt, Phys. Rev. {\bf 160}, 1113 (1967).
\bibitem{juan} J. F. Barbero and J. P\'erez-Mercader, Phys. Rev. D
{\bf 48}, 3663 (1993).
\bibitem{adm} R. Arnowitt, S. Deser and C. W. Misner,
{\it The dynamics of
general relativity}, in {\it Gravitation} (ed. L.Witten, Wiley, New
York, 1962), p. 227-265.
\bibitem{kt} E. W. Kolb and M. S. Turner, in {\it The Early Universe}
(Addison-Wesley, 1990), p. 64.
\bibitem{cern} F. Dydak, in {\it Proc. 25th Int. Conf. on High Energy
Physics} (ed. Rhua, World Scientific, Singapore, 1991).
\bibitem{olive94} K. A. Olive and G. Steigman, astro-ph/9405022,
preprint (1994).
\bibitem{wea} T. P. Walker, G. Steigman, D. N. Schramm, K. A. Olive
and H. Kang, Ap.J. {\bf 376}, 51 (1991).
\bibitem{grm} P. M. Jeffrey and E. Anders, Geochim. Cosmochim. Acta
{\bf34}, 1175 (1970).
\bibitem{sw} J. Geiss, P. Eberhardt, F. Buhler, J. Meister and
P. Singer, J. Geophys. Res. {\bf 75}, 5972 (1970).
\bibitem{ls} D. C. Black, Geochim. Cosmochim. Acta {\bf36}, 347 (1972).
\bibitem{7Li1} J. Yang, M. S. Turner, G. Steigman, D. N. Schramm and
K. A. Olive, Ap.J. {\bf 281}, 493 (1984).
\bibitem{7Li2} K. A. Olive, D. N. Schramm, G. Steigman and
T. P. Walker, Phys. Lett. B, {\bf 236}, 454 (1990).
\bibitem{cc1} P. Eberhardt, in {\it Proc. 9th Lunar Planet. Sci.
Conf.}, 1027 (1978).
\bibitem{cc2} U. Frick and R. K. Moniot, in {\it Proc. 8th Lunar Planet.
Sci. Conf.}, 229 (1977).
\bibitem{oea} K. A. Olive, D. N. Schramm, G. Steigman and
T. P. Walker, Phys.
Lett. B {\bf 236}, 454 (1990).
\bibitem{neutron} W. Mampe, P. Ageron, C. Bates, J. M. Pendebury and
A. Steyerl, Phys. Rev. Lett. {\bf 63A}, 593 (1989).
\bibitem{kernan1} P. Kernan, OSU Physics Ph. D. Thesis (1993).
\bibitem{kernan2} P. Kernan and L. Krauss, Preprint astro-ph/9402010,
{\it to appear in} Phys. Rev. Lett..
\bibitem{tabla} {\it Review of Particle Properties}, Phys. Rev. D {\bf
45}, Part 2 (June 1992).
\bibitem{giulini} D. Giulini and C. Kiefer, Preprint gr-qc/9405040.


\end{thebibliography}
\end{document}